\documentclass[iop,twocolappendix,numberedappendix,appendixfloats]{emulateapj}
\usepackage{amsmath}

\newcommand{\be}{\begin{equation}}
\newcommand{\ee}{\end{equation}}
\newcommand{\bea}{\begin{eqnarray}}
\newcommand{\eea}{\end{eqnarray}}

\def\teff{T_{\rm eff}}
\def\logg{{\rm log}\,g}

\shortauthors{CONROY ET AL.}

\shorttitle{They Might Be Giants}

\begin{document}


\title{They Might Be Giants: An efficient color-based selection of red
  giant stars}

\author{Charlie Conroy, Ana Bonaca, Rohan P. Naidu, Daniel J. Eisenstein,
  Benjamin D. Johnson, Aaron Dotter, Douglas P. Finkbeiner}

\affil{Harvard-Smithsonian Center for Astrophysics, Cambridge, MA 02138, USA}

\slugcomment{Submitted to ApJ}

\begin{abstract}

  We present a color-based method for identifying red giants based on
  Pan-STARRS $grz$ and {\it WISE} $W1$ and $W2$ photometry.  We
  utilize a subsample of bright stars with precise parallaxes from
  Gaia DR2 to verify that the color-based selection reliably separates
  dwarfs from giants.  The selection is conservative in the sense that
  contamination is small ($\approx30$\%) but not all giants are
  included (the selection primarily identifies K giants).  The
  color-based selection can be applied to stars brighter than
  $W1\approx16$, more than two magnitudes fainter than techniques
  relying on shallower 2MASS photometry.  Many streams and clouds are
  visible in the resulting sky maps, especially when binned by Gaia
  DR2 proper motions, including the Sagittarius stream, the
  Hercules-Aquila Cloud, the Eastern Banded Structure, Monoceros, and
  the Virgo Overdensity. In addition to the characterization of new
  and known stellar streams, we expect that this method for selecting
  red giants will enable detailed analysis of the diffuse stellar halo
  to distances exceeding 100 kpc.

\end{abstract}



\section{Introduction}
\label{s:intro}

The stellar halo bears witness to the assembly of our Galaxy.  The
distribution of halo stars on small scales is highly structured,
indicating that the majority have been accreted from satellite
galaxies \citep[e.g.,][]{Bell08}, as expected in the cold dark matter
cosmological paradigm \citep[e.g.,][]{White78, Bullock05}.  In the
inner $\sim30\,$kpc, remnants of individual dwarf galaxies and
globular clusters have been discovered as overdensities of old and
metal-poor turn-off stars \citep[][and references therein]{Newberg16,
  Grillmair16}, and more recently via kinematics \citep[e.g.,][]{Malhan18}.

The virial radius of the Milky Way extends beyond 250\,kpc
\citep[e.g.,][]{Posti18}, but due to the scarcity of luminous tracers,
the outer halo remains poorly charted.  RR Lyrae, pulsating standard
candles, have been used to map the largest volume of the Galactic
halo, tracing the smooth component out to $\sim110\,$kpc
\citep{Cohen17}, and remnants of disrupted dwarf galaxies between 10
and 100\,kpc \citep[e.g.,][]{Vivas06, Watkins09, Sesar17b}.  In
addition to requiring time series observations to detect RR Lyrae,
they are also relatively rare, making it difficult to use them to
trace lower mass populations \citep{Sesar14}.  Other rare, luminous
tracers have been used to map the outer halo including blue horizontal
branch stars \citep[BHB][]{Deason12, Deason18b} and Carbon stars
\citep{Mauron04, Mauron08}. A full account of the Milky Way's
accretion history will require mapping the halo with tracers that are
both luminous and abundant.

Red giant branch stars are both relatively numerous and luminous and
therefore would serve as an excellent tracer of the stellar halo.  The
key challenge is identifying them in photometric surveys.  Red giants
have similar optical colors as the much more numerous dwarfs of the
same temperature; however, they can be separated with near-infrared
photometry \citep[e.g.,][]{Bessell88}.  NIR color-based selection was
used to identify M giants and map not only the massive debris from the
Sagittarius dwarf galaxy \citep[e.g.,][]{Majewski03}, but also stellar
streams and clouds from less massive progenitors
\citep[e.g.,][]{RochaPinto03,RochaPinto04}, and identify very distant
halo stars \citep{Bochanski14}.  \citet{Koposov15} and \citet{Li16}
refined this selection method and detected parts of the Sagittarius
stream in the highly crowded and high extinction plane of the Galaxy.

Previous work utilizing NIR color-based selection has focused on
identifying M giants, which are both intrinsically luminous and
clearly localized in 2MASS and {\it WISE} color-color diagrams.  An
advantage of this approach is that both 2MASS and {\it WISE} are
all-sky surveys and so structure can be mapped in this way throughout
the Galaxy.  The disadvantages are two-fold: M giants are a relatively
rare population, compared to e.g., K giants, and the use of 2MASS data
restricts the usable data to $K_s<13.5$ (equivalent to $W1<13.5$ for
cool stars).  {\it WISE} data extend at least 2.5 mag fainter than
2MASS, so a color-based method for selecting red giants that does not
require 2MASS photometry would enable a view of structure in the
Galaxy to much fainter limits than previous 2MASS-based catalogs.  In
this paper, we present such a method for a pure selection of (mostly
K-type) giant stars relying only on Pan-STARRS and {\it WISE}
photometry.


\begin{figure*}[!t]
\center
\includegraphics[width=0.95\textwidth]{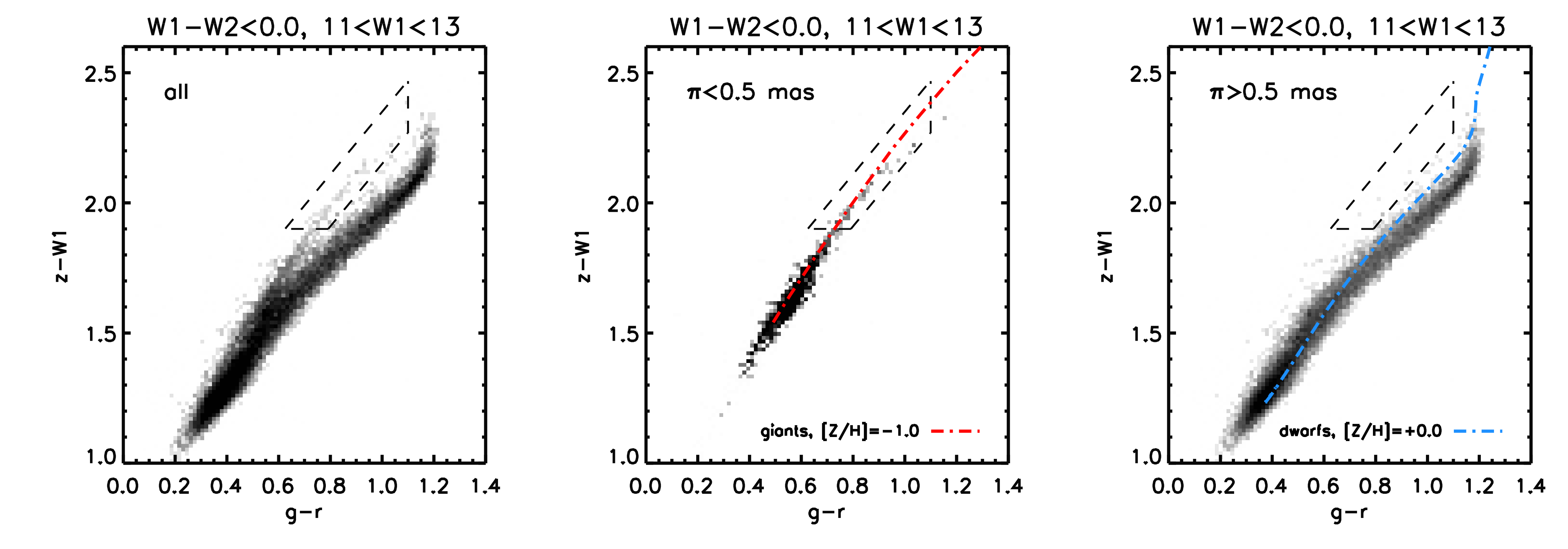}
\vspace{0.1cm}
\caption{Selection of giants via color-color cuts.  Each panel shows
  log(number) of stars in the (de-reddened) $g-r$ vs. $z-W1$ color
  space selected to have $W1-W2<0$, $11<W1<13$, and
  $80^\circ<b<90^\circ$.  Two sequences are clearly seen in the left
  panel, which shows all stars.  The middle and right panels show
  stars selected to have parallaxes $<0.5$ mas (middle) and $>0.5$ mas
  (right).  Our selection box used to identify giants is marked by the
  dashed lines.  Also shown in the middle and right panels are the
  giant (log$\,g<3$) and dwarf (log$\,g>4$) sequences from a 10 Gyr
  MIST isochrone.}
\label{fig:ccut}
\end{figure*}

\section{Selecting Giants}
\label{s:data}

We begin with a catalog of stars that is cross-matched between the
Gaia mission \citep{GC16}, data release 2 \citep{GC18a}, Pan-STARRS
data release 1 \citep[PS1][]{Chambers16}, and {\it WISE} $W1$ and $W2$
photometry \citep{Wright10, Cutri13}.  The cross-matching was
performed using the Large Survey Database framework \citep{Juric12}
with a matching radius of $<1\arcsec$.  All photometry has been
corrected for Galactic extinction using the \citet{Schlegel98} dust
maps.  Where Gaia data are used we require
\texttt{visibility\_periods\_used}$\ge6$ and
\texttt{astrometric\_excess\_noise}$\ge1.2\gamma(G)$ where
$\gamma(G)={\rm max}[1,10^{0.2(G-18)}]$ \citep[see][for
details]{Lindegren18}.  When {\it WISE} data are used we require
uncertainties on $W1$ and $W2$ photometry to be $0<\sigma_{W1}<0.1$
and $\sigma_{W2}>0$.  Following \citet{Li16}, we also apply the
following criteria to ensure high-quality photometry:
\texttt{ext\_flag}$=$0 and \texttt{cc\_flags}$=$`00'.

It is well known that a single color is generally unable to separate
the K and M dwarfs from the K and M giants.  This ambiguity presents a
critical bottleneck to studying the stellar halo because along any
line of sight a flux-limited sample will be overwhelmingly dominated
by the much more numerous dwarfs.  A combination of broadband colors
that could reliably separate the dwarfs and giants would enable a much
cleaner view of the outer regions of the Galaxy.  Previous work along
these lines have used 2MASS $JHK_s$ photometry \citep{Majewski03}, or
a combination of 2MASS and {\it WISE} photometry \citep{Koposov15} to
select M giants.

Inspired by previous efforts, we explored a variety of color-color
cuts utilizing Pan-STARRS and {\it WISE} broadband colors.  After some
experimentation we settled on the following selection: 
\be
\begin{split}
-0.4< W1- W2<0.0\\
g-r<1.1\\
1.9<z-W1<2.5\\
1.2(g-r)+0.95 < z- W1< 1.2(g-r)+1.15
\end{split}
\ee
\noindent
The $W1-W2$ selection produces a very clear bifurcation in $g-r$
vs. $z-W1$ which we identify as sequences of dwarfs and giants (see
Figure \ref{fig:ccut}).  The additional cuts isolate the giant
sequence in that space.  We chose to avoid the use of 2MASS photometry
because 2MASS is much shallower than {\it WISE} and our goal is to go
as faint as possible.  The fairly conservative $z-W1>1.9$ selection
was motivated by the larger photometric scatter at fainter magnitudes.
The depth in our case is limited by {\it WISE}: at $W1=16$ the typical
uncertainty on $W1$ is 0.06 mag and at $W2=15$ the typical uncertainty
on $W2$ is 0.07 mag.

An illustration of the selection method is presented in Figure
\ref{fig:ccut}.  We select stars with Galactic latitude $b>80^\circ$
to minimize the effects of reddening and $11<W1<13$ so that
uncertainties on the Gaia DR2 parallaxes are small.  We then apply the
$W1-W2<0.0$ cut and plot the remaining stars in the left panel.  The
selection box in the $g-r$ vs. $z-W1$ space is indicated by the dashed
lines.  In the middle panel we show only stars with a parallax of
$\pi<0.5$ mas ($>2$ kpc), and in the right panel those stars with
$\pi>0.5$ mas ($<2$ kpc).  70\% of the stars in the selection box have
$\pi<0.5$ mas.  As we will see in a moment, stars with low parallaxes
are giants, indicating that our color selection has a purity of $\approx70$\%.

In the middle and right panels we also include the giant (log$\,g<3$)
and dwarf (log$\,g>4$) sequences from MIST isochrones at 10 Gyr
\citep{Choi16}.  Stars in the middle panel are similar to
low-metallicity giants, as expected for a halo population, while stars
in the right panel are similar to solar metallicity dwarfs, as
expected for the local disk population.  The models are systematically
bluer in $g-r$ for the reddest colors; this is a known limitation of
the color-temperature relations used in the models \citep{Choi16}.

According to the MIST isochrones, the color selection is identifying
the middle portion of the RGB (e.g., K giants) over the metallicity
range $-2<$[Z/H]$<0$.  The coolest stars (M giants) are omitted due to
the $g-r<1.1$ cut.  This cut was necessary in order to avoid
substantial contamination from the M dwarf locus, which turns vertical
in Figure \ref{fig:ccut} at $g-r>1.1$.  The absolute magnitudes of the
RGB stars range from $-5\lesssim M_{\rm W1}\lesssim-2$.  The {\it
  WISE} data reaches $W1\approx15.5$ before photometric uncertainties
compromise the color selection.  This translates into a reach of
$>100$ kpc for this color selection technique.

\begin{figure}[!t]
\center
\includegraphics[width=0.45\textwidth]{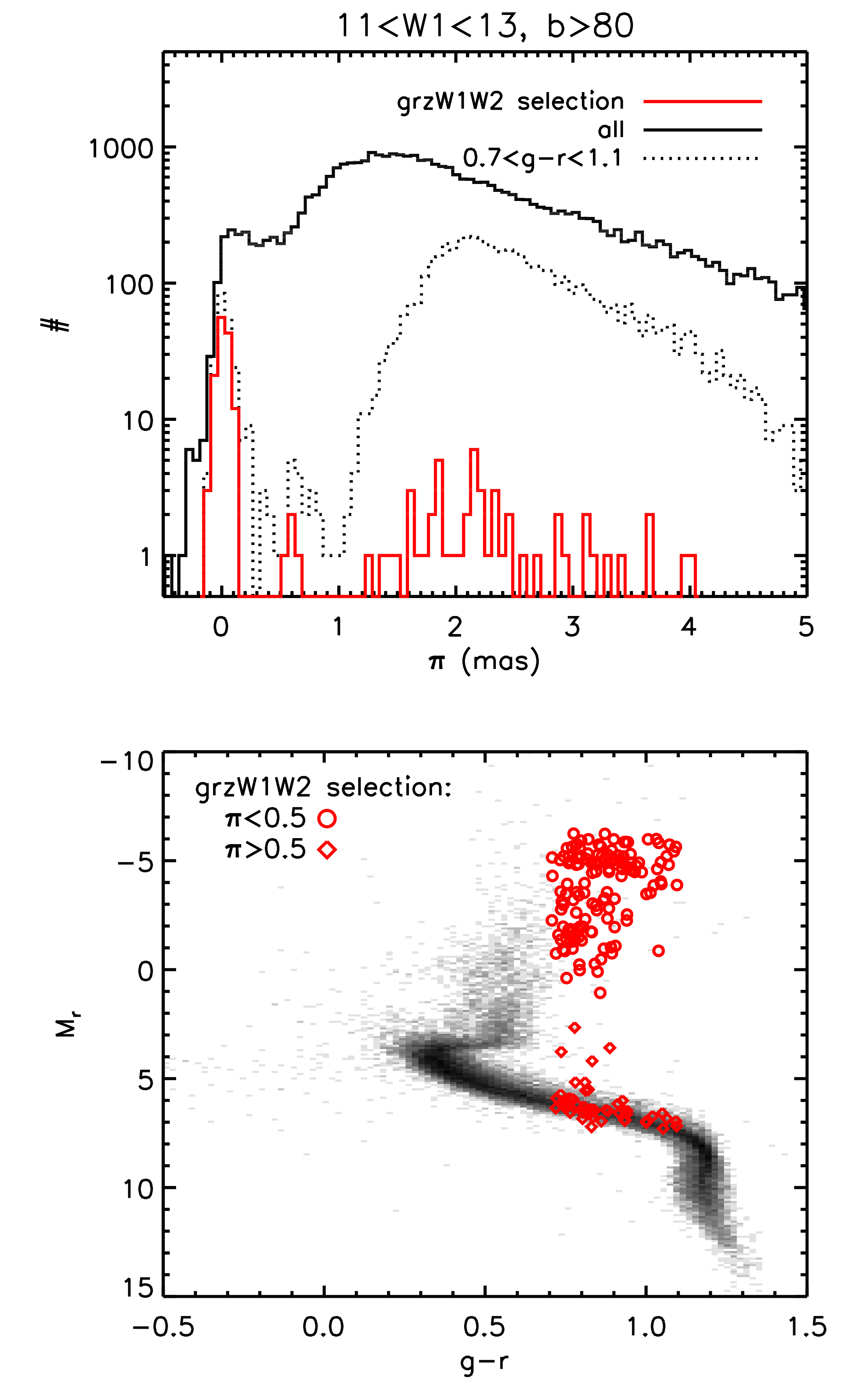}
\vspace{0.1cm}
\caption{{\it Top panel:} Distribution of parallaxes for stars with
  $11<W1<13$ and $b>80^\circ$.  The entire population is compared to
  stars with $0.7<g-r<1.1$, and our giant-based color selection
  (labeled as $grzW1W2$).  These bright stars have small parallax
  uncertainties and so the stars with $\pi\approx0.0$ can be
  confidently associated with greater distances.  The greater distance
  combined with the narrow magnitude range suggests that the
  low-parallax stars are giants.  This is confirmed in the bottom
  panel, which shows the color-magnitude diagram for all stars (shown
  as a Hess diagram with a logarithmic color stretch) and for the
  giant-based color selection.  Stars with $\pi<0.5$ mas are clearly
  giants, and they comprise 70\% of the $grzW1W2$ color-selected
  stars. Stars with negative parallaxes were assigned a nominal
  parallax of 0.01 mas for display purposes.}
\label{fig:sel2}
\end{figure}

In Figure \ref{fig:sel2} we examine in more detail the purity of the
proposed color selection.  In the top panel we show the distribution
of parallaxes for a sample of stars with $11<W1<13$ and $b>80^\circ$.
The overall sample is compared to subsamples defined via a simple
color cut of $0.7<g-r<1.1$, and our $grzW1W2$ color selection.  This
bright sample of stars has small parallax uncertainties and so one
sees a clear bimodality in the distribution of parallaxes.  This is
due to the large difference in luminosities between dwarfs and giants
combined with the relatively narrow magnitude range considered in the
figure.  The bottom panel shows the associated color-magnitude
diagram, where it is clear that stars with $\pi<0.5$ mas are red
giants.  Returning to the top panel, comparison of the dotted and red
solid lines highlights the enormous suppression of dwarfs provided by
our $grzW1W2$ color selection compared to a simple $g-r$ selection.

We explore the completeness of these color cuts using the SDSS SEGUE
sample \citep{Yanny09}, which includes stellar parameters
\citep{LeeYS08}.  We restrict the SEGUE sample in several respects,
including $|b|>20^\circ$, $W1<15.5$, and requiring a quality flag of
`nnnnn'.  Following \citet{Xue14}, we identify K giants with
$0.5<g-r<1.3$.  We furthermore require [Fe/H]$>-2$, as some stars have
what appear to be unrealistically low metallicities (e.g., some stars
have [Fe/H]$\approx-4$).  We selected giants based on the clear
separation from the dwarf sequence in $\logg-\teff$ space.  We
discovered that some of the resulting K giant stars had $\pi>0.5$ mas,
which, given their apparent magnitudes, places them squarely on the
photometric dwarf sequence.  Such stars are removed.  We find that our
color cuts select 80\% of the SEGUE K giants with $\teff<4500$K.  In
other words, our selection method identifies cooler K giants with high
completeness.  The completeness drops rapidly for warmer giants, which
is not surprising given the selection of red stars with both $g-r$ and
$z-W1$ cuts.

\begin{figure}[!t]
\center
\includegraphics[width=0.45\textwidth]{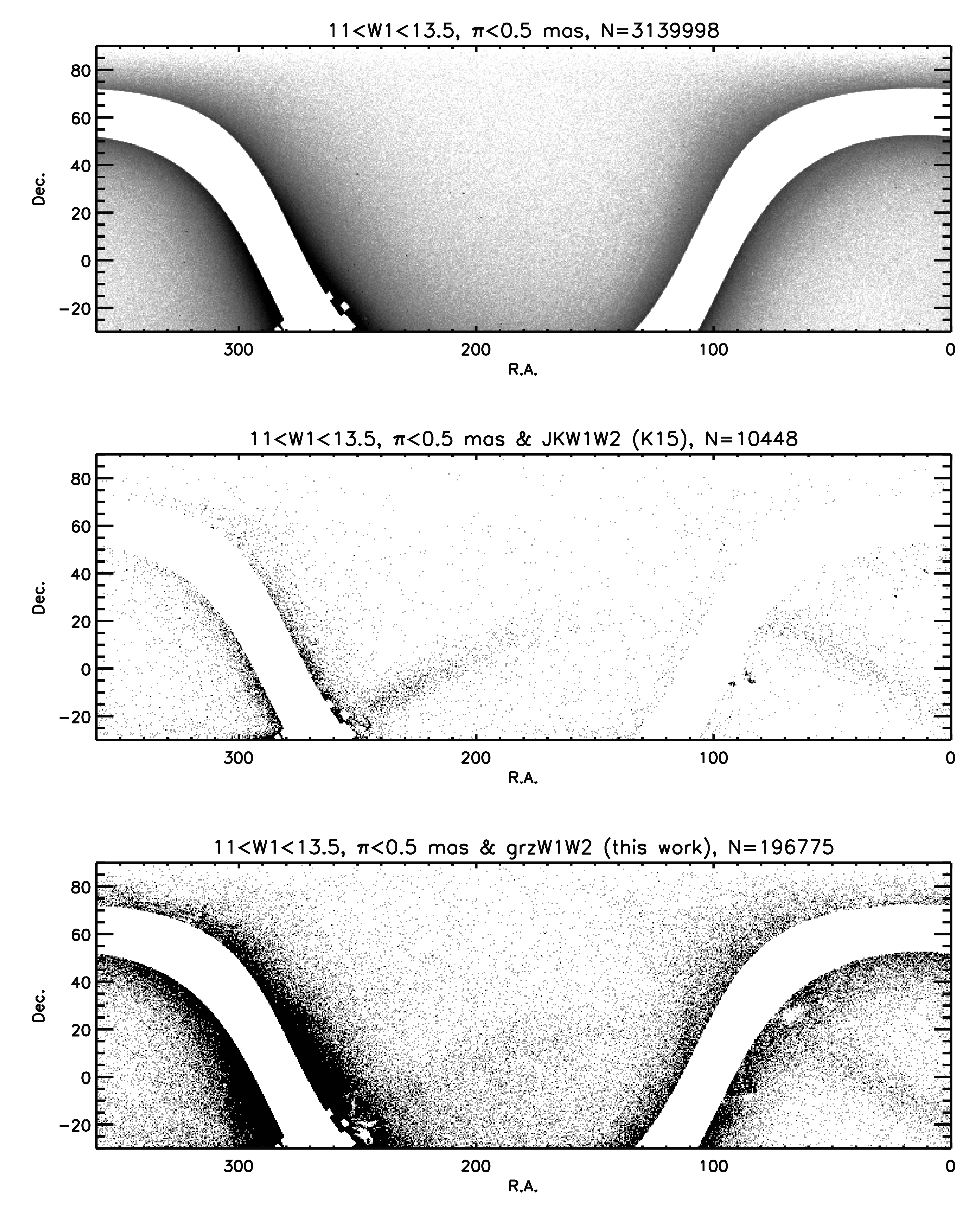}
\vspace{0.1cm}
\caption{Comparison between parallax-only selection (top panel),
  parallax plus M giant selection from \citet{Koposov15} (middle
  panel), and parallax plus our K giant color selection (bottom
  panel), for stars with $11<W1<13.5$.  The maps show log(number) of
  stars in $0.5^\circ\times0.5^\circ$ bins.  The color stretch is the
  same for the middle and lower panel. The total number of stars in
  each map is shown in the title of each panel.}
\label{fig:parcomp}
\end{figure}

\begin{figure*}[!t]
\center
\includegraphics[width=0.98\textwidth]{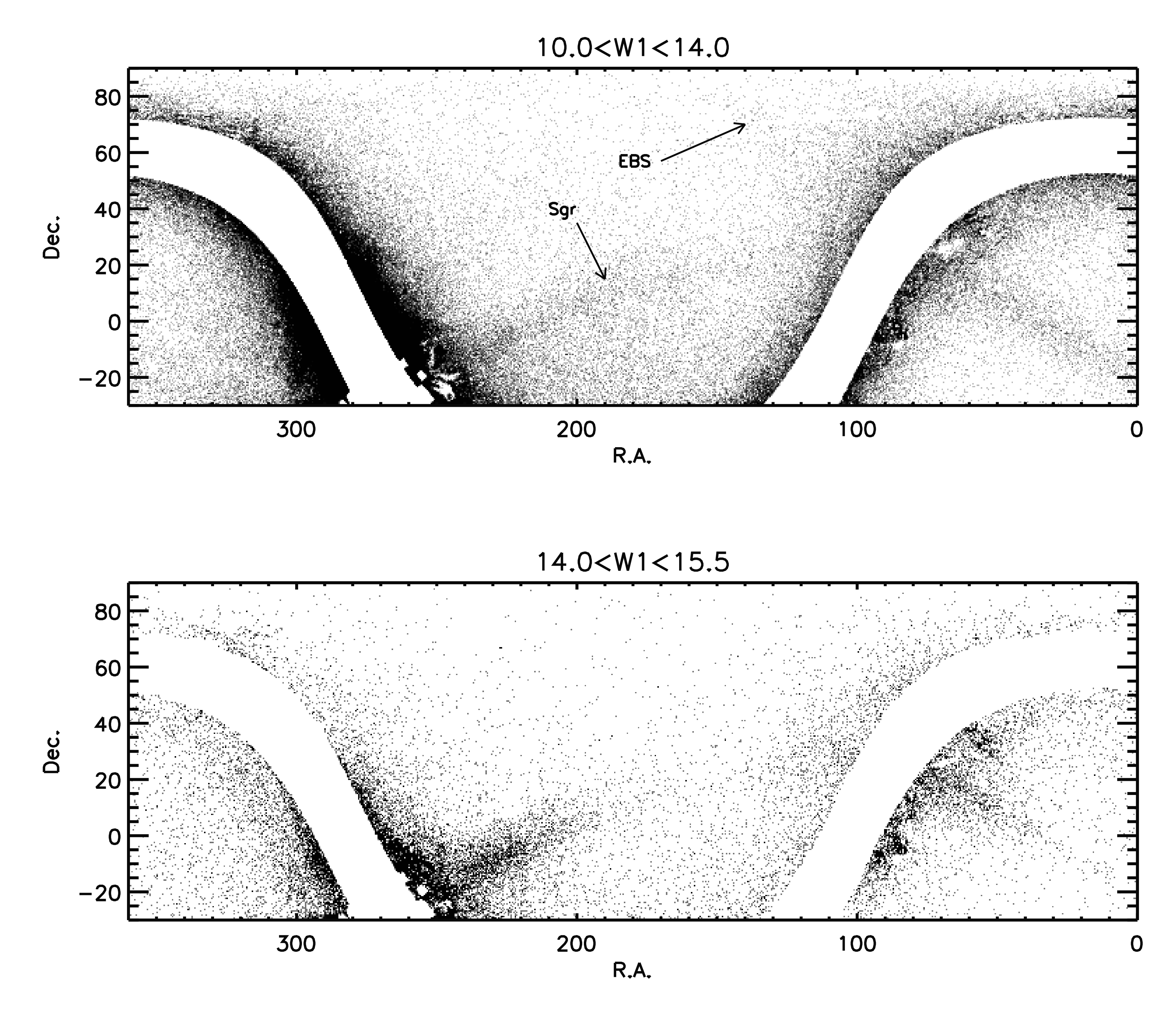}
\vspace{0.1cm}
\caption{Maps of red giant stars selected according to the color cuts
  in Figure \ref{fig:ccut}.  Regions where $|b|<10^\circ$ are omitted.
  The panels show stars in two $W1$ magnitude ranges.  The maps show
  log(number) of stars in $0.5^\circ\times0.5^\circ$ bins.  The
  Sagittarius stream (labeled) is prominent in the top and middle
  panels.  The Eastern Banded Structure (EBS) is visible and labeled
  in the top panel. }
\label{fig:map}
\end{figure*}


\section{Results}
\label{s:res_map}

We applied the color-based selection of red giants to the entire
cross-matched Gaia DR2, PS1, and {\it WISE} catalogs.  The sky coverage of
PS1 limits us to Dec$>-30^\circ$.  We have also removed stars with
$\pi>0.5$ mas as an additional filter against foreground stars.

Figure \ref{fig:parcomp} compares binned maps of stars selected only
by parallax (top panel) and those with both a parallax and two
color-based selection techniques (middle and bottom panels).  Here we
include stars with $11<W1<13.5$ and omit data near the plane
($|b|<10^\circ$).  In the middle panel we implement the
\citet{Koposov15} M giant selection, which utilizes 2MASS $JK_s$ and
{\it WISE} $W1$ and $W2$ photometry (these authors restricted their
analysis to $11<W1<13.5$, which motivated our choice for the magnitude
interval for this figure).  In the bottom panel we use our new color
selection.  It is clear that even for these relatively bright stars,
where the mean parallax uncertainty is 0.05 mas, that a parallax
selection alone is insufficient to identify the distant giants.  This
is not surprising --- a star at 10 kpc has a true parallax of 0.1 mas
and so a typical parallax uncertainty (for the stars in this figure)
of 0.05 mas implies that it will be difficult to separate a star at 10
kpc from the much more numerous foreground dwarfs.  In contrast, the
parallax plus color-based selection reveals a variety of structures
and overdensities, the most prominent being the Sagittarius (Sgr)
stream.

\begin{figure*}[!t]
\center
\includegraphics[width=0.95\textwidth]{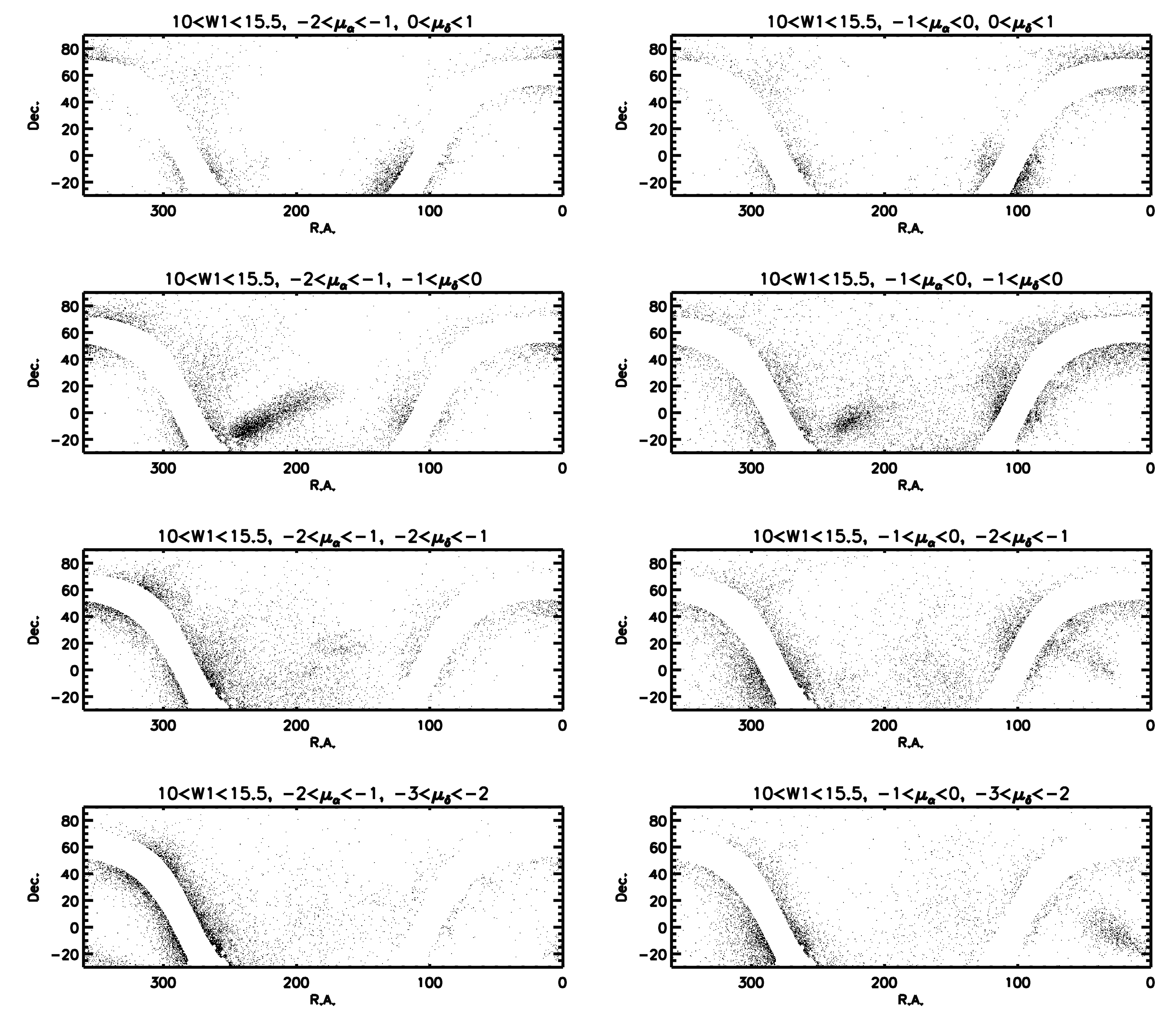}
\vspace{0.1cm}
\caption{Maps of red giant stars in a wide magnitude range
  ($10<W1<15.5$) in bins of proper motion.  Rows are sorted by
  increasing $\mu_\delta$ while columns are sorted by increasing
  $\mu_\alpha$ (in units of mas yr$^{-1}$).  The maps show the number
  of stars in $0.5^\circ\times0.5^\circ$ bins.  See Figure
  \ref{fig:pm2} for a continuation of the proper motion bins.  Many
  known streams and stellar overdensities appear in these proper
  motion maps.  }
\label{fig:pm1}
\end{figure*}

\begin{figure*}[!t]
\center
\includegraphics[width=0.95\textwidth]{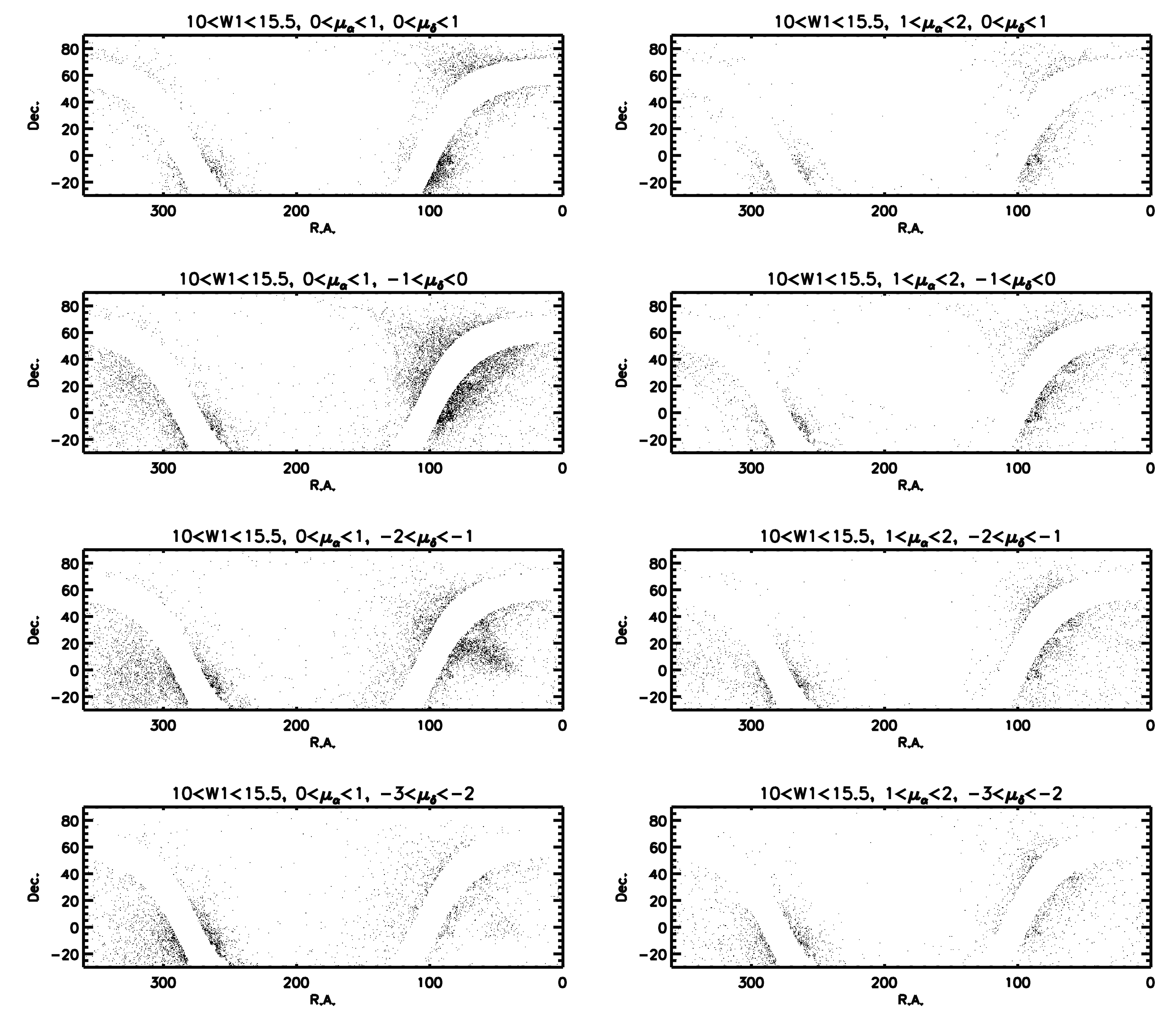}
\vspace{0.1cm}
\caption{Continuation of Figure \ref{fig:pm1}.}
\label{fig:pm2}
\end{figure*}

Comparison of the middle and bottom panels reveals the differences
between an M giant-based selection (middle) and a K giant based
selection (bottom).  There are approximately $26\times$ more stars in
the lower panel compared to the middle panel, even though the
magnitude range is the same, owing largely to the fact that K giants
are much more numerous than M giants.  Because of this, one can see
more streams and structures in the lower panel.  As we will see below,
a second advantage of the Pan-STARRS and {\it WISE} color selection
employed here is that we can extend to fainter limits than selections
based on 2MASS and {\it WISE}.  We do note that an advantage of the
color selection in the middle panel is that it can be applied to the
entire sky thanks to the all-sky 2MASS and {\it WISE} datasets.

In Figure \ref{fig:map} we show maps of the red giant stars in two
$W1$ magnitude bins.  Data in the plane ($|b|<10^\circ$) have been
omitted.  These maps are rich in structure.  The most obvious feature
is the Sgr stream which stretches across the entire map in the top
panel.  In the middle panel the Sgr stream has broken up into pieces,
and in the bottom panel Sgr is not easily visible, except perhaps for
the overdensity near (260,-10); see below for details.  At these
fainter magnitudes we are likely probing the more distant components
of the Sgr stream, as seen in previous work \citep{Majewski03} and
predicted by models \citep[e.g.,][]{Law10, Dierickx17}.

In the top panel there are several additional easily visible features
including the stream extending from (100,40) to (160,90).  This
structure is at an approximately constant Galactic latitude of
$b\approx35^\circ$ and corresponds to Feature B in \citet{Slater14},
also known as the Eastern Banded Structure reported in
\citet{Grillmair11}.

In Figures \ref{fig:pm1} and \ref{fig:pm2} we show the giant star maps
in proper motion bins.  In these maps the structures in our Galaxy
appear most dramatic.  In addition to the Sgr stream, which appears in
many panels \citep[see][for previous proper motion measurements at
various locations along the Sgr stream using {\it Hubble Space
  Telescope}]{Sohn15}, one clearly sees the structure referred to as
either a part of the Monoceros Ring \citep{Slater14} or the Eastern
Banded Structure \citep{Grillmair11} in the $0<\mu_\alpha<1$,
$-1<\mu_\delta<1$ bins.  This feature extends from (100,40) to
(180,90).  \citet{Deason18a} used SDSS-Gaia proper motion measurements
to demonstrate that this structure is part of a complex network of
sub-structures in the Galactic anti-center region with a likely origin
due to some perturbation of the Galactic disk.

The Hercules-Aquila Cloud \citep{Belokurov07, Simion14, Simion18}
appears at $290<$R.A.$<360$ and $-30<$Dec$<40$ in the proper motion
range $0<\mu_\alpha<1$, $-3<\mu_\delta<0$.  The plume of stars in the
bin $-2<\mu_\alpha<-1$, $-1<\mu_\delta<0$, north of Sgr, may also be
associated with Hercules-Aquila.  The proper motion gradient of this
structure, extending thousands of sq. degrees across the sky, is
remarkable and supports the scenario outlined in \citet{Simion18} that
the Hercules-Aquila Cloud is part of a much larger debris structure
originating from an old, well-mixed accretion event.

There is a large cloud centered at (150,-10) in the proper motion bin
$-1<\mu_\alpha<0$, $-2<\mu_\delta<-1$, with a plausible northward
extension in the bin $-1<\mu_\alpha<0$, $-3<\mu_\delta<-2$.  This
structure is very likely associated with the Virgo Overdensity
\citep{Newberg02, Bonaca12, Duffau14, Vivas16, Sesar17}.  If so, then
the map in Figure \ref{fig:pm1} offers the most complete on-sky
extension of the Virgo Overdensity to-date.

There are additional features in these proper motion maps whose
association with known structures is less obvious.  We leave a
detailed analysis of the structure in these diagrams to future work.


\section{Summary}
\label{s:sum}

In this paper we have presented a new color-based method for selecting
K-type red giants that is based on Pan-STARRS $grz$ and {\it WISE}
$W1$ and $W2$ photometry.  Gaia DR2 parallaxes of a bright subsample
confirms that this selection identifies giants with low contamination
from dwarfs ($\approx30$\%).  Comparison to the SEGUE sample of
spectroscopically-confirmed giants reveals that the completeness of
our color selection is $\approx80$\% for cool ($\teff<4500$K) giants.
The resulting maps display a rich variety of structure both as a
function of R.A. and Dec. and proper motion.

Our proposed color selection offers several benefits over previous NIR
color-based techniques that focused on selection of M giants
\citep[e.g.,][]{Majewski03, Koposov15}.  First, M giants, though more
luminous, are rarer than K giants, and so a selection aimed at
identifying the latter class of objects will result in a higher
density of tracers.  Moreover, previous M giant selections relied on
2MASS photometry, which is approximately 2.5 mag shallower than {\it
  WISE}.  By avoiding 2MASS we are therefore able to reach at least 2
mag deeper than previous giant-based color selections.  The major
drawback to our approach is that it requires $grz$ photometry, for
which complete coverage exists only at Dec$>-30^\circ$.

In the future we will investigate and characterize the many features
visible in the maps.  These maps can also be used to study the diffuse
stellar halo at large distances.  The requirement that these stars be
detected in {\it WISE} $W1$ and $W2$ means that they are relatively
bright, and so will be straightforward to follow up with high
resolution spectroscopy on $6-10$m telescopes.  Pan-STARRS limited the
sky coverage to Dec$>-30^\circ$ but the technique can be easily
extended to the south with Dark Energy Survey data.  Furthermore, the
overall depth can be extended by utilizing the extended {\it WISE} 4
yr data.

We emphasize that the color cuts adopted herein are not necessarily
optimal for selecting red giants as simple trial and error was used to
arrive at the final selection.  The success of our adopted selection
suggests that more sophisticated techniques \citep[e.g.,][]{Mints17,
  Anderson18} should have even greater success at identifying red
giants, so long as NIR data are employed.


\acknowledgments 

We thank the referee for a prompt and constructive report.  CC
acknowledges support from the Packard Foundation.  RPN acknowledges
support from a James Mill Peirce Fellowship and an Ashford Graduate
Fellowship.  DJE is supported as a Simons Foundation Investigator.
This work has made use of data from the European Space Agency (ESA)
mission Gaia (https://www.cosmos.esa.int/gaia), processed by the Gaia
Data Processing and Analysis Consortium (DPAC,
https://www.cosmos.esa.int/web/gaia/dpac/consortium). Funding for the
DPAC has been provided by national institutions, in particular the
institutions participating in the Gaia Multilateral Agreement.




\end{document}